\documentclass{mn}

\usepackage{epsfig}
\usepackage{latexsym,subfigure}

\def \Halpha {${\rm H} \alpha $} 
\def \paperone  {Norberg \etal (2001)}

\def \etapar {$\eta $}
\def \bj {b$_{\rm J}$}

\def \zmin {$z_{\rm min}$}
\def \zmax {$z_{\rm max}$}

\def \sqdeg {$\Box^{\circ}$}

\def \Mstar {M$^\star$\,} 

\def \bj {\rm b$_{\rm J}$}

\def \mpc {$h^{-1} {\rm{Mpc}}$}

\def \and   {\rm {et al.} \rm}  
\def \etal  {\rm {et al.} \rm}

\begin{document}

\title[The 2dF Galaxy Redshift Survey: 
The dependence of galaxy clustering on luminosity and spectral type.
]
{
\vspace{-0.75cm}The 2dF Galaxy Redshift Survey: The dependence of galaxy 
clustering on luminosity and spectral type.}
\vspace{-0.5cm}

\author[The 2dFGRS collaboration]{
\parbox[t]{\textwidth}{
\vspace{-1.0cm}
Peder Norberg$^{1}$,
Carlton M.\ Baugh$^{1}$,
Ed Hawkins$^{2}$,
Steve Maddox$^{2}$,
Darren Madgwick$^{3}$, %
Ofer Lahav$^{3}$, 
Shaun Cole$^{1}$, 
Carlos S.\ Frenk$^{1}$, 
Ivan Baldry$^{4}$, %
Joss Bland-Hawthorn$^{5}$,
Terry Bridges$^{5}$, 
Russell Cannon$^{5}$, 
Matthew Colless$^{6}$, 
Chris Collins$^{7}$, 
Warrick Couch$^{8}$, 
Gavin Dalton$^{9}$,
Roberto De Propris$^{8}$, 
Simon P.\ Driver$^{10}$, 
George Efstathiou$^{3}$, 
Richard S.\ Ellis$^{11}$, 
Karl Glazebrook$^{4}$, 
Carole Jackson$^{6}$,
Ian Lewis$^{5}$, 
Stuart Lumsden$^{12}$, 
John A.\ Peacock$^{13}$,
Bruce A.\ Peterson$^{6}$, 
Will Sutherland$^{13,9}$,
Keith Taylor$^{11,5}$
}
\vspace*{6pt} \\ 
$^{1}$Department of Physics, University of Durham, South Road, 
    Durham DH1 3LE, UK \\ 
$^{2}$School of Physics \& Astronomy, University of Nottingham,
       Nottingham NG7 2RD, UK \\
$^{3}$Institute of Astronomy, University of Cambridge, Madingley Road,
    Cambridge CB3 0HA, UK \\
$^{4}$Department of Physics \& Astronomy, Johns Hopkins University,
       Baltimore, MD 21218-2686, USA \\
$^{5}$Anglo-Australian Observatory, P.O.\ Box 296, Epping, NSW 2121,
    Australia\\  
$^{6}$Research School of Astronomy \& Astrophysics, The Australian 
    National University, Weston Creek, ACT 2611, Australia \\
$^{7}$Astrophysics Research Institute, Liverpool John Moores University,  
    Twelve Quays House, Birkenhead, L14 1LD, UK \\
$^{8}$Department of Astrophysics, University of New South Wales, Sydney, 
    NSW 2052, Australia \\
$^{9}$Department of Physics, University of Oxford, Keble Road, 
    Oxford OX1 3RH, UK \\
$^{10}$School of Physics and Astronomy, University of St Andrews, 
    North Haugh, St Andrews, Fife, KY6 9SS, UK \\
$^{11}$Department of Astronomy, California Institute of Technology, 
    Pasadena, CA 91125, USA \\
$^{12}$Department of Physics, University of Leeds, Woodhouse Lane,
       Leeds, LS2 9JT, UK \\
$^{13}$Institute for Astronomy, University of Edinburgh, Royal Observatory, 
       Blackford Hill, Edinburgh EH9 3HJ, UK \\
\vspace*{-0.5cm}}

\maketitle 
 
\begin{abstract}
We investigate the dependence of galaxy clustering on luminosity
and spectral type using the 2dF Galaxy Redshift Survey (2dFGRS).
Spectral types are assigned using the principal component analysis
of Madgwick et al. We divide the sample into two broad spectral
classes: galaxies with strong emission lines (`late-types'), and
more quiescent galaxies (`early-types'). We measure the clustering
in real space, free from any distortion of the clustering pattern
due to peculiar velocities, for a series of volume-limited
samples. The projected correlation functions of both spectral types
are well described by a power law for transverse separations in the
range 2$<$($\sigma/h^{-1}$\,Mpc)$<$15, with a marginally steeper
slope for early-types than late-types. Both early and late types
have approximately the same dependence of clustering strength on
luminosity, with the clustering amplitude increasing by a factor of
$\sim$2.5 between $L^*$ and 4$L^*$. At all luminosities, however,
the correlation function amplitude for the early-types is $\sim$50\%
higher than that of the late-types. These results support the view
that luminosity, and not type, is the dominant factor in determining
how the clustering strength of the whole galaxy population varies 
with luminosity.
\end{abstract}

\begin{keywords}
methods: statistical - methods: numerical - 
large-scale structure of Universe - galaxies: formation
\vspace*{-1.25cm}
\end{keywords}

\section{Introduction} 
\label{sec:intro}

One of the major goals of large redshift surveys like the 
2 degree Field Galaxy Redshift Survey (2dFGRS) is to 
make an accurate measurement of the spatial distribution 
of galaxies. The unprecedented size of 
the 2dFGRS makes it possible to quantify how the clustering 
signal depends on intrinsic galaxy properties, such as luminosity or 
star formation rate.

The motivation behind such a program is to characterize the 
galaxy population and to provide constraints 
upon theoretical models of structure formation. 
In the current paradigm, galaxies form in dark matter 
haloes that are built up in a hierarchical way through 
mergers or by the accretion of smaller objects. 
The clustering pattern of galaxies is therefore determined by 
two processes: the spatial distribution of dark matter haloes  
and the manner in which dark matter haloes are populated by 
galaxies (Benson \etal 2000b; Peacock \& Smith 2000; Seljak 2000; 
Berlind \& Weinberg 2002). 
The evolution of clumping in the dark matter has been studied 
extensively using N-body simulations of the growth of 
density fluctuations via gravitational instability (e.g. 
Jenkins \etal 1998; 2001).
With the development of powerful theoretical tools that can follow 
the formation and evolution of galaxies in the hierarchical scenario,  
the issue of how galaxies are apportioned amongst dark matter 
haloes can be addressed, and detailed predictions of the 
clustering of galaxies are now possible 
(Kauffmann, Nusser \& Steinmetz 1997; Kauffmann \etal 1999; 
Benson \etal 2000a,b; Somerville \etal 2001). 

The first attempt to quantify the difference between the clustering 
of early and late-type galaxies was made using a shallow 
angular survey, the Uppsala catalogue, with morphological types 
assigned from visual examination of the 
photographic plates (Davis \& Geller 1976). 
Elliptical galaxies were found to have a higher amplitude angular 
correlation function than spiral galaxies. In addition, the 
slope of the correlation function of ellipticals was steeper than 
that of spiral galaxies at small angular separations. 
More recently, the comparison of clustering for different types 
has been extended to three dimensions using redshift surveys.
Again, similar conclusions have been reached in these studies, 
namely that ellipticals have a stronger clustering amplitude 
than spirals (Lahav \& Saslaw 1992; Santiago \& Strauss 1992; 
Iovino \etal 1993; Hermit \etal 1996; Loveday \etal 1995; 
Guzzo \etal 1997; Willmer \etal 1998).

The subjective process of visual classification can now be superseded by   
objective, automated algorithms to quantify the shape of a galaxy. 
One recent example of such a scheme can be found in Zehavi \etal (2002), 
who measured a ``concentration parameter'' for $30\,000$ galaxy images 
from the Sloan Digital Sky Survey, derived from the radii of different 
isophotes.
Again, based upon cuts in the distribution of concentration parameter, 
early-types are found to be more clustered than late-types.

In this paper, we employ a different method to classify galaxies, 
based upon a principal component analysis (PCA) of galaxy spectra, 
which is better suited to the 2dFGRS data (Madgwick \etal 2002). 
This technique has a number of attractive features.
First, the PCA approach is completely objective and reproducible. 
An equivalent analysis can, 
for example, be applied readily to spectra produced by theoretical models of 
galaxy formation or to spectra obtained in an independent redshift survey. 
Secondly, the PCA can be applied over the full magnitude range of the 2dFGRS, 
whenever the spectra are of sufficient signal to noise (see 
Section~\ref{subsec:eta_class}). For the 2dFGRS, the image quality 
is adequate to permit a visual determination of 
morphological type only for galaxies brighter than $b_{\rm J}\,\simeq\,17$, 
which comprise a mere $5\%$ of the spectroscopic sample. 

Two previous clustering studies have used spectral information 
to select galaxy samples. Loveday, Tresse \& Maddox (1999) 
grouped galaxies in the Stromlo-APM redshift survey into three classes 
based upon the equivalent width of either the \Halpha\ or OII lines, and 
found that galaxies with prominent emission lines display weaker 
clustering than more quiescent galaxies.
Tegmark \& Bromley (1999) measured the relative bias between different 
spectral classes in the Las Campanas redshift survey (Shectman \etal 1996), 
using a classification derived from PCA analysis (Bromley \etal 1998), and 
also found that early spectral types are more strongly clustered than 
late spectral types.  (See also Blanton 2000 for a revision of 
Tegmark \& Bromley's analysis, which takes into account the effect 
of errors in the survey selection function.) 

Here, we use the 2dFGRS survey to measure the dependence of 
galaxy clustering jointly on luminosity and spectral type, adding an 
extra dimension to the analysis carried out by \paperone. Previously, a 
pioneering study of bivariate galaxy clustering, in terms of luminosity 
and morphological type, was carried out using the Stromlo-APM 
redshift survey (Loveday \etal 1995). To place the analysis presented  
here in context, the samples that we consider cover a larger volume 
and, despite being volume-limited (see Section~\ref{sec:vol.lim}), 
typically contain over an order of magnitude more galaxies 
than those available to Loveday \etal 

We give a brief overview of the 2dFGRS in Section~\ref{sec:data}, along 
with details of the spectral classification and an explanation of how 
the samples used in the clustering analysis were constructed. 
The estimation of the redshift space correlation function and its real 
space counterpart, the projected correlation function, are outlined in 
Section~\ref{sec:meth}. A brief overview of the clustering 
of 2dFGRS galaxies in redshift space, selected by luminosity and spectral 
type, is given in Section~\ref{sec:redspace}; a more detailed analysis of 
the redshift space clustering can be found in Hawkins \etal (2002).
We present the main results of the paper in Section~\ref{sec:res} and 
conclude in Section~\ref{sec:conc}.

\section{The data}
\label{sec:data}

\subsection{The 2dFGRS sample}

Detailed descriptions of the construction of the 2dFGRS and 
its properties are given by Colless \etal (2001).
In summary, galaxies are selected down to a magnitude limit of 
$b_{\rm J}\approx\,19.45$ from the APM Galaxy Survey 
(Maddox \etal 1990a,b, 1996, 2002).
The sample considered in this paper consists of over $160\,000$ redshifts 
measured prior to May 2001.
We focus our attention on the two large contiguous volumes of the 
survey, one centred on the Southern Galactic Pole (hereafter SGP) and 
the other close to the direction of the Northern Galactic Pole (NGP).

\subsection{Spectral classification of 2dFGRS galaxies}
\label{subsec:eta_class}
\begin{figure}
\epsfig{file=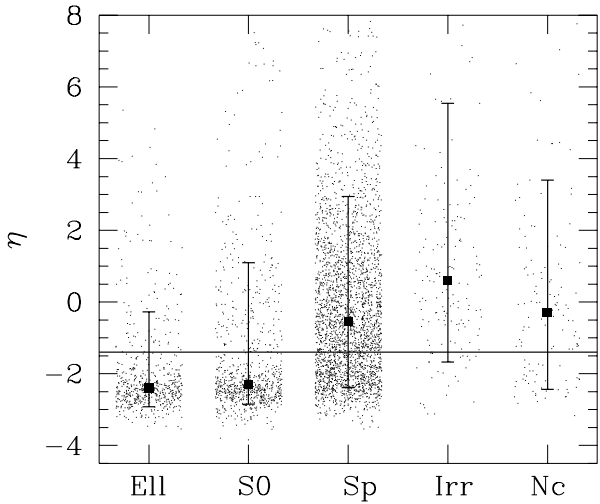,width=0.47\textwidth,clip=,bbllx=40,bblly=180,bburx=480,bbury=545}
\caption{
A comparison between the morphological classification of 
bright (\bj$\,<\,$17.0) APM galaxies  by Loveday (1996) with the 
2dFGRS spectral classification, as quantified by the continuous variable 
\etapar\ 
(see text and Madgwick \etal 2002 for the definition). 
The morphological classification distinguishes between 
elliptical (Ell), lenticular (S0), spiral (Sp) and irregular (Irr) galaxies.
All galaxies with both a morphological classification and a spectral 
classification are plotted. The Non-classified (Nc) class contains  
objects for which morphological classification was attempted but for which 
Loveday was unable to assign a type.
The squares show the median value of \etapar\ for each morphological class 
defined by Loveday, and the error bars show the $10$-$90$ percentile range.
}
\label{fig:jon.eta}
\end{figure}

\begin{figure}
\epsfig{file=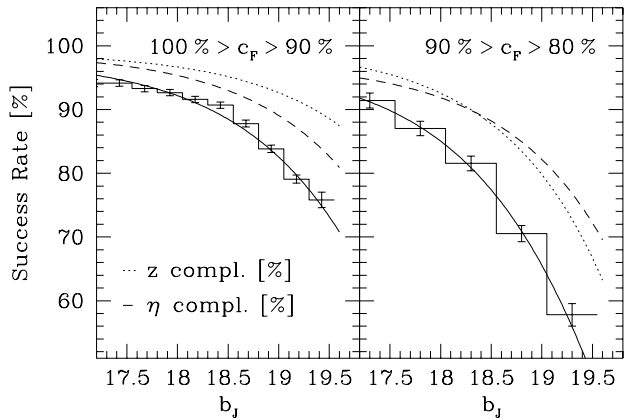,width=0.47\textwidth,clip=,bbllx=50,bblly=370,bburx=550,bbury=710}
\caption{
The histograms, plotted with Poisson error bars, show the 
success rate for assigning a spectral type to a targeted 
galaxy as a function of apparent magnitude. Two field 
completeness ($c_{\rm F}$, defined in text) ranges are shown, 
as indicated by the values at the top of each panel.
The redshift completeness, {\it i.e.} the fraction of 
targeted galaxies for which a redshift is measured, 
is shown by the dotted lines.
The spectral classification completeness, {\it i.e.} 
the fraction of galaxies with measured redshifts below $z=0.15$ 
that have spectra of sufficient signal-to-noise to be used in the PCA,
is shown by the dashed lines.
(The dotted and dashed curves show parametric fits to the inferred 
redshift completeness and spectral classification completeness 
respectively.)
The model for the spectral classification success rate, 
shown by the solid lines, 
is the product of the dotted and dashed lines in each panel, and is 
a good fit to the histogram in each case.
}
\label{fig:eta.compl}
\end{figure}

The spectral properties of 2dFGRS galaxies are characterized using 
the principal component analysis (PCA) described by Madgwick \etal (2002).
This analysis makes use of the spectral information in the 
rest-frame wavelength
range 3700\AA\ to 6650\AA, thereby including all the major optical
diagnostics between OII and \Halpha.  For galaxies with $z>0.15$,
sky absorption bands contaminate the \Halpha\ line.  
Since this can affect the stability of the classification, we restrict 
our analysis to galaxies with $z<0.15$ following Madgwick et al. (2002).

The 2dFGRS spectra are classified by a single parameter, \etapar, which is a
linear combination of the first and second principal components.  This
combination has been chosen specifically to isolate the relative
strength of emission and absorption lines present in each galaxy's spectrum,
thereby providing a diagnostic which is robust to the instrumental
uncertainties that affect the calibration of the continuum.
Physically, this parameter is related to the star formation rate in 
a galaxy, as is apparent from the tight correlation of \etapar\  
with the equivalent width of \Halpha\ in emission line 
galaxies (Bland-Hawthorn \etal 2002). 
In this paper, we divide the 2dFGRS sample into two broad, distinct classes: 
galaxies with spectra for which the PCA returns $\eta<-1.4$, 
and which we refer to, for the sake of brevity, as early-type, and 
galaxies with $\eta>-1.4$, which we call late-type.
The distribution of \etapar\ for 2dFGRS spectra displays a shoulder 
feature at this value (see Fig.~4 of Madgwick \etal 2002).

The spectral type of a galaxy, as given by the value of \etapar, 
clearly depends upon its physical properties and is therefore a useful 
and effective way in which to label galaxies.
Nevertheless, it is still instructive to see how well, if at all, 
\etapar\ correlates with the morphogical type assigned 
in a subjective fashion from a galaxy image.
Madgwick \etal (2002) show that there is a reasonable correspondence 
between \etapar\ and morphological type, using high signal-to-noise 
spectra and photometry taken from Kennicutt (1992); $\eta\simeq-1.4$ 
approximately delineates the transition between early and late 
morphological types in \bj. 
We revisit the comparison between classifications based on 
spectral and morphological types in Fig.~\ref{fig:jon.eta}, 
this time using 2dFGRS spectra and UK Schmidt images.
The horizontal axis shows the morphological type assigned to 
a subset of bright APM galaxies by Loveday (1996).
Although there is a substantial amount of scatter in the \etapar\ 
values of spectra that lie within a given morphological 
class, it is reassuring to see that the median \etapar\ does correlate 
with morphological class. Moreover, the median \etapar\ values match 
up well with the broad division that we employ to separate early 
and late types. Galaxies denoted ``early-type'' on the basis of 
their morphology have a median \etapar\ that is smaller than our 
fiducial value of $\eta=-1.4$ and vice-versa for late-types.
In practice, for the samples analysed in this paper, the correspondence 
between morphological type and spectral class  will be better than 
suggested by Fig.~\ref{fig:jon.eta}.
This is because the sample used in the comparison in Fig.~\ref{fig:jon.eta} 
consists of nearby extended galaxies, and so the distribution of spectral 
types is distorted somewhat by aperture effects 
(see e.g. Kochanek, Pahre \& Falco 2002; Madgwick et al. 2002).
This effect arises because the fibres used to collect the galaxy
spectra are of finite size (subtending 2'' on the sky).  For this reason, 
when we measure the spectrum of a nearby galaxy it is possible that a 
disproportionate amount of light will be sampled from the bulge, thereby making
the galaxy appear systematically redder or ``earlier" in type.  We find
that this effect is only significant for the most nearby galaxies 
($z < 0.05$) and should be completely negligible beyond $z\sim0.1$ 
(Madgwick et al. 2002).

\subsection{Sample Selection}
\label{sec:compl}

In order to construct an optimal sample for the measurement of  
the two point correlation function, we select regions 
with high completeness in terms of measured redshifts, using a redshift 
completeness mask for the 2dFGRS, similar to the one described in 
Colless \etal (2001; see also Norberg \etal 2002). 
Such a mask is required because of the tiling strategy adopted to 
make the best use of the allocated telescope time, along with the fact 
that the survey is not yet finished. 
An additional consideration is the success rate with which spectral types 
have been assigned to galaxies, which depends upon the signal-to-noise 
ratio of the galaxy spectrum. 

In Fig.~\ref{fig:eta.compl}, we show histograms of the spectral 
classification success rate for two different ranges of field 
completeness, $c_{\rm F}$, which is defined as the ratio 
of the number of measured redshifts in a given 2dF field to the 
number of targets.
The spectral classification success rate has two contributions. 
The first of these is the redshift completeness, shown by the dotted curve. 
This incompleteness arises because we do not always succeed in measuring 
a redshift for a targeted object.
The redshift incompleteness is necessarily small for the high completeness 
fields contributing to the histograms.
The second contribution is the spectral classification completeness. 
Galaxies do not recieve a spectral classification when a redshift is measured 
with $z\le0.15$ (and is therefore within the redshift range over 
which the PCA can be carried out), but the spectrum has too small 
a signal-to-noise ratio for the PCA to be applied successfully  
(typically ${\rm S}/{\rm N}<10$). 
The spectral classification success rate is given by the 
product of these two contributions. Our model for this effect, 
plotted as the solid curves in each panel of Fig.~\ref{fig:eta.compl}, 
is in good agreement with the success rate realised in the 2dFGRS, 
shown by the histograms. 

Rather than weight the data to compensate for a spectral classification 
success rate below $100\%$, we instead modulate the number of unclustered or 
random points laid down in each field in the clustering analysis 
to take into account the varying success rate.
We have conducted a number of tests in which we varied the completeness 
thresholds used, adopted different weighting schemes using samples 
of higher completeness, and we have also compared our results with those 
from Norberg \etal (2001), whose samples are not subject to 
spectral classification 
incompleteness. The results of these tests confirm 
that our clustering measurements are robust to changes 
to the details of how we model the incompleteness; 
this is largely due to our practice of restricting the 
analysis to high completeness fields.
Excluding areas below our relatively high sector completeness 
threshold (see Colless \etal 2001 for a definition), 
we estimate that the effective solid angle used in the SGP region 
is $\sim$ 380 \sqdeg, and in the NGP 250 \sqdeg.

\begin{figure*}
\begin{center}
\vspace*{-1.0cm}
\subfigure[$-18.0\,\ge\,M_{b_{\rm J}}-5\log_{10}\,h\,\ge\,-19.0$]{\epsfig{file=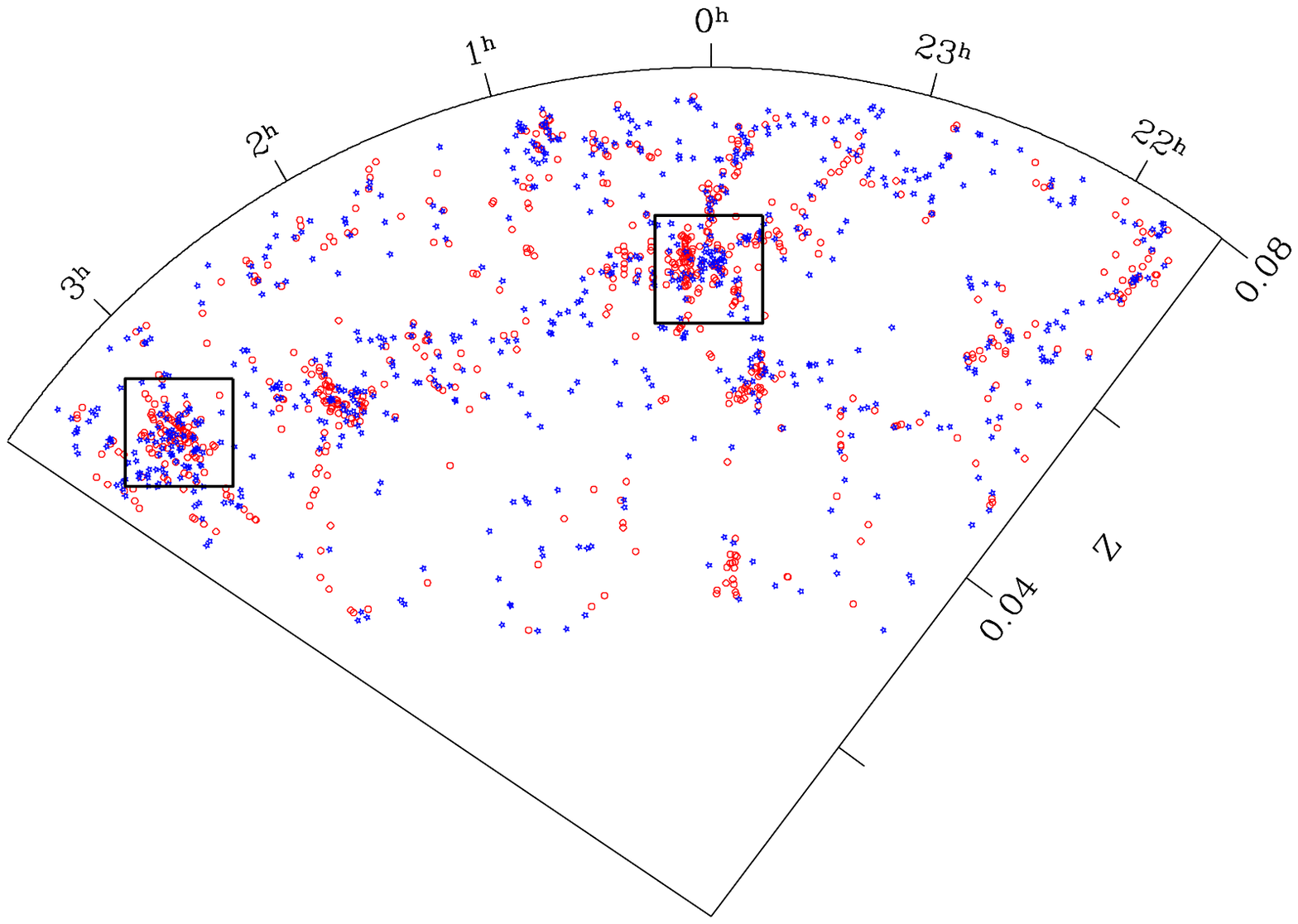,width=0.57\textwidth,clip=,angle=90,bbllx=315,bblly=85,bburx=573,bbury=545}}\vspace*{-0.5cm}
\subfigure[$-20.0\,\ge\,M_{b_{\rm J}}-5\log_{10}\,h\,\ge\,-21.0$]{\epsfig{file=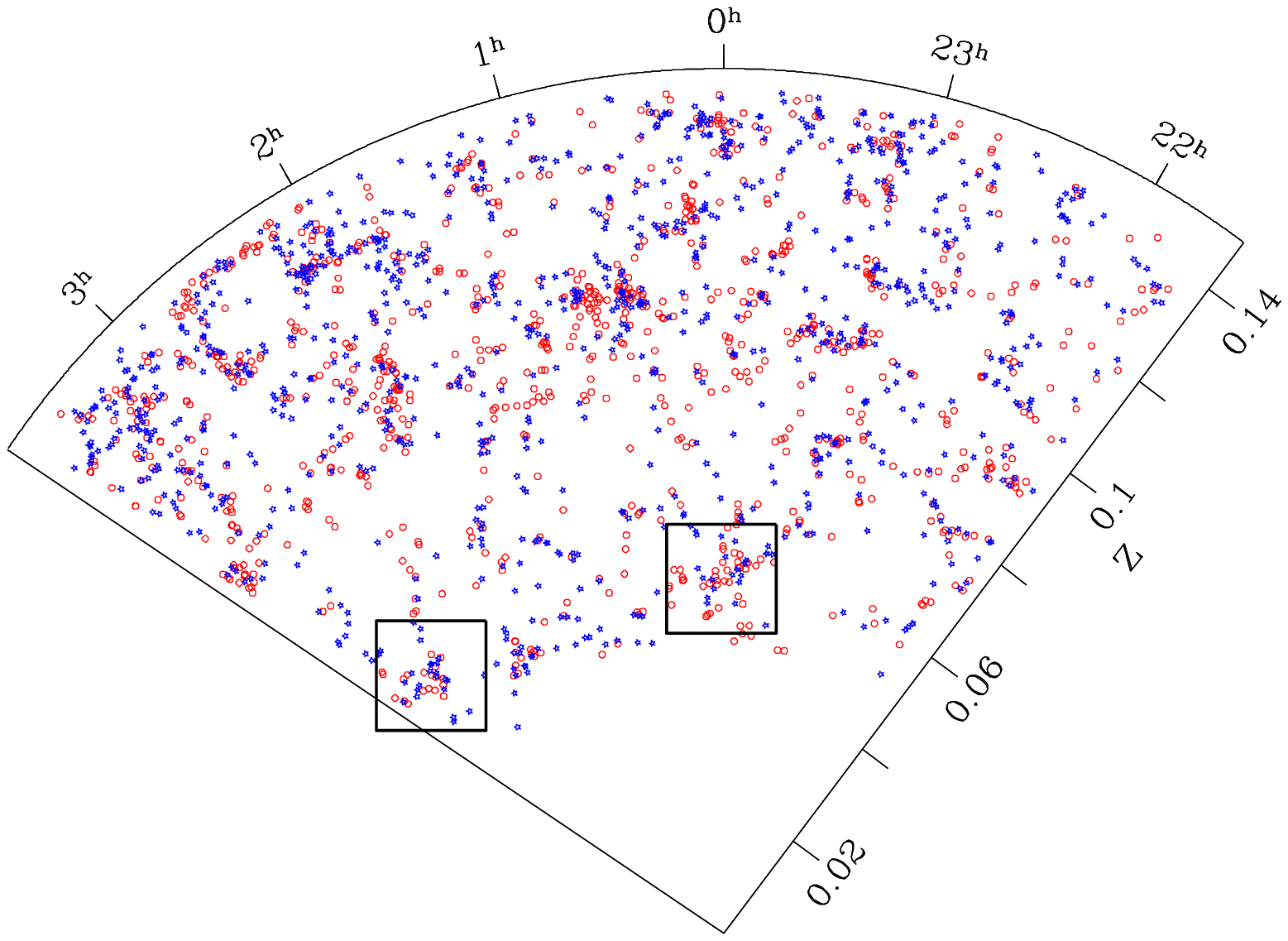,width=0.57\textwidth,clip=,angle=90,bbllx=305,bblly=85,bburx=563,bbury=545}}
\caption{
The spatial distribution of 2dFGRS galaxies in the SGP. 
The panels show the redshift and right ascension of galaxies in a three 
degree thick strip in declination for different magnitude ranges.
To expand the scale of the plot, the redshift range shown has been restricted; 
note that the redshift scales are different in the two panels. 
Blue stars mark the locations of late-type galaxies and red circles the 
positions of early-type galaxies. The boxes show the two structures which 
are referred to in the text.
}
\label{fig:coneplot}
\end{center}
\end{figure*}

\subsection{Constructing a volume-limited sample}
\label{sec:vol.lim}

We analyse a series of volume-limited samples 
drawn from the 2dFGRS, following the strategy adopted by \paperone. 
The chief advantage of this approach is simplicity; the radial 
distribution of galaxies is uniform apart from modulations in  
space density due to clustering.
Therefore, the complication of modelling the radial selection 
function in a flux-limited survey is avoided. This 
is particularly appealing for the current analysis, as separate 
selection functions would be required for each class of spectral 
type studied, since Madgwick \etal (2002) have demonstrated 
that galaxies with different spectral types have different 
luminosity functions. 

The disadvantage of using volume-limited samples is that 
a large fraction of galaxies in the flux-limited 
catalogue do not satisify the selection criteria. 
As \paperone\ point out, a volume-limited sample specified by a 
range in absolute magnitude has both a lower (\zmin) and an upper 
redshift cut (\zmax), because the flux-limited catalogue has, in 
practice, bright and faint apparent magnitude limits.
This seemingly profligate use of galaxy redshifts was a serious problem for  
previous generations of redshifts surveys. This is not, however, the case   
for the 2dFGRS, which contains sufficient galaxies to permit the 
construction of large volume-limited samples defined both by luminosity 
and spectral type. As we demonstrate in section~\ref{sec:res}, 
the volume-limited samples we analyse are large enough, both in 
terms of volume and number of galaxies, to give extremely 
robust clustering measurements.

To construct a volume-limited sample, it is necessary to estimate 
the absolute magnitude that each galaxy would have at $z=0$. 
This requires assumptions about the variation of galaxy luminosity 
with wavelength and with redshift, or equivalently, with 
cosmic time. We make use of the class dependent $k$-corrections 
derived by Madgwick \etal (2002). The mean weighted 
$k$-corrections are given by the following expressions:
\begin{eqnarray}
k(z) &=& 2.6 z + 4.3 z^2 \,\,\, ({\rm early-types})  \\ 
k(z) &=& 1.5 z + 2.1 z^2 \,\,\, ({\rm late-types}) \\
k(z) &=& 1.9 z + 2.7 z^2 \,\,\, ({\rm full\,\,sample}).  
\end{eqnarray}
These $k$-corrections have the appeal that they are extracted directly 
in a self-consistent way from 2dFGRS spectra. However, no 
account is taken of evolution in the galaxy spectrum. 
The explicit inclusion of evolution could 
lead to the ambiguous situation whereby a galaxy's spectral 
type changes with redshift. 
We have checked that our results are, in fact, insensitive 
to the precise choice of $k$-correction, comparing clustering 
results obtained with the spectral type dependent $k$-corrections 
given above with those obtained when a global $k+e$-correction 
(i.e. making an explicit attempt to account for galaxy evolution, 
albeit in an average sense) is applied  (as in Norberg \etal 2001).

Since the $k$-corrections are class dependent, the 
\zmin\ and \zmax\ values corresponding to a given absolute magnitude 
range are also slightly class dependent. Hence, the volumes defining 
the samples for two different spectral classes for the same bin in 
absolute magnitude will not coincide exactly.
In addition to this subtle class dependent definition of the volumes, the 
values of \zmin\ and \zmax\ vary slightly with position on the sky. 
This is due to revisions made to the map of galactic extinction 
(Schlegel, Finkbeiner \& Davis 1998) and to the CCD recalibration of APM 
plate zero-points since the definition of the original input catalogue. 

Finally, throughout the paper, we adopt an $\Omega_{0}=0.3$, 
$\Lambda_{0}=0.7$ cosmology to convert redshift into comoving distance. 
The relative clustering strength of our samples is insensitive to this 
choice.

\section{Estimating the two-point correlation function}
\label{sec:meth}

The galaxy correlation function is estimated on a two dimensional 
grid of pair separation parallel ($\pi$) and perpendicular ($\sigma$) 
to the line-of-sight. 
To estimate the mean density of galaxy pairs, a catalogue of randomly 
positioned points is generated with the same angular distribution and 
the same values of \zmin\ and \zmax\ as the data, taking into 
account the completeness of the survey as a function of position on 
the sky, as described in Section~\ref{sec:compl}.
The correlation function is estimated using 
\begin{equation}
\xi_{\rm{H}} = \frac{DD\,RR}{DR^2}\,-\,1\,   , 
\label{eq:ham}
\end{equation}
where $DD$, $DR$ and $RR$ are the numbers of  
data-data, data-random and random-random pairs respectively in 
each bin (Hamilton 1993).
This estimator does not require an explicit estimate of 
the mean galaxy density.
We have also cross-checked our results using the estimator 
proposed by Landy \& Szalay (1993): 
\begin{equation}
\xi_{\rm{LS}} = \frac{ DD - 2DR + RR }{RR},
\label{eq:ls} 
\end{equation}
where, this time $DD$, $DR$ and $RR$ are the suitably normalised 
numbers of data-data, data-random and random-random pairs.
We find that the two estimators give the same results over the 
range of pair separations in which we are interested.

The clustering pattern of galaxies is distorted when radial positions 
are inferred from redshifts, as expected in the gravitational instability 
theory of structure formation (e.g. Kaiser 1987; 
Cole, Fisher \& Weinberg 1994). 
Clear evidence for this effect is seen in the shape of the 
two point correlation function when plotted as $\xi(\sigma,\pi)$, 
as demonstrated clearly for galaxies in the 2dFGRS   
by Peacock \etal (2001) and for groups of galaxies in the Zwicky 
catalogue by Padilla \etal (2001).
After giving a brief flavour of the clustering of 2dFGRS galaxies 
in redshift space in Section~\ref{sec:redspace}, we focus our 
attention on clustering in real space in the remainder of the paper.
The clustering signal in real space is inferred by 
integrating $\xi(\sigma,\pi)$ 
in the $\pi$ direction (i.e. along the line of sight): 
\begin{equation} 
\frac{\Xi(\sigma)}{\sigma} = \frac{1}{\sigma} \int_{-\infty}^{\infty} 
\xi(\sigma,\pi) {\rm d}\pi.
\label{eq:chiofsigma} 
\end{equation}
For the samples that we consider, the integral converges  
by pair separations of $\pi\,\ge\,50\,$\mpc.
The projected correlation function can then be written as an 
integral over the spherically averaged real space correlation 
function, $\xi(r)$,
\begin{equation} 
\frac{\Xi(\sigma)}{\sigma} = \frac{2}{\sigma} \int_{\sigma}^{\infty} \xi(r) 
\frac{r{\rm d}r}{\left(r^{2}-\sigma^{2}\right)^{1/2}}, 
\label{eq:projxi}
\end{equation}
(Davis \& Peebles 1983). 
If we assume that the real space correlation function is a power 
law (which is a fair approximation for APM galaxies out to separations 
around $r\sim10\,h^{-1}\,$Mpc, see e.g. Baugh 1996), then 
Eq.~\ref{eq:projxi} can be written as 
\begin{equation} 
\frac{\Xi(\sigma)}{\sigma} = \left(\frac{r_{0}}{\sigma} \right)^{\gamma}
\frac{\Gamma(1/2)\Gamma( [\gamma-1]/2)}
{\Gamma(\gamma/2)} = \left(\frac{r_{0}}{\sigma} \right)^{\gamma} A(\gamma), 
\label{eq:pow}
\end{equation}
where $\Gamma(x)$ is the usual Gamma function, 
and we have used $\xi(r) = (r_{0}/r)^{\gamma}$, where $r_{0}$ is the real 
space correlation length and $\gamma$ is equal to the slope of the 
projected correlation function $\Xi(\sigma)/\sigma$.
As we demonstrate in Section~\ref{sec:xi.s}, the projected correlation 
function is well described by a power law. 

We study a range of samples containing different numbers of 
galaxies and covering different volumes of the Universe.
It is imperative to include sampling fluctuations 
when estimating the errors on the measured correlation function, 
to allow a meaningful comparison of the results obtained from 
different samples. This contribution to the errors has  
often been neglected in previous work.
Following \paperone, we employ a sample of 22 mock 2dFGRS catalogues 
drawn from the $\Lambda$CDM Hubble Volume simulation (Evrard \etal 2002) 
to estimate the error bars on the measured correlation functions. 
The construction of these mock catalogues is explained in 
Baugh \etal (2002, in preparation; see also Cole \etal 1998 and 
Norberg \etal 2002). 
These catalogues have the same selection criteria and the same 
clustering amplitude as measured for galaxies in the flux-limited 2dFGRS.
We have experimented with ensembles of mock catalogues constructed 
with different clustering strengths to ascertain how best to assign error 
bars when the measured clustering has a different amplitude from that of our 
fiducial sample of 22 2dFGRS mocks. 
We found that the error bars obtained directly by 
averaging over a test ensemble of mocks are reproduced most closely by 
using the scaled fractional {\it rms} scatter 
derived from the fiducial ensemble of 22 mocks, rather 
than by taking the absolute error.

\section{Clustering in Redshift Space}
\label{sec:redspace}
In this section we give a brief overview of the clustering 
of 2dFGRS galaxies in redshift space, for samples selected 
by luminosity and spectral type. 
First, in Section~\ref{sec:cone}, we give a qualitative impression 
of the clustering differences by plotting the spatial distribution of 
galaxies in volume-limited samples. 
Then we quantify these differences by measuring the spherically 
averaged correlation function, $\xi(s)$.
A more comprehensive analysis of the clustering of 2dFGRS galaxies 
in redshift space will be presented by Hawkins \etal (2002).

\subsection{Spatial distribution of 2dFGRS galaxies}
\label{sec:cone}

It is instructive to gain a visual impression of the spatial 
distribution of 2dFGRS galaxies before interpreting the measured 
correlation functions. 
In Fig.~\ref{fig:coneplot},  we show the spatial distribution of 
galaxies in two ranges of absolute magnitude: in panel (a) we show 
a sample of faint galaxies 
($-18.0\,\ge\,M_{b_{\rm J}}-5\log_{10}\,h\,\ge\,-19.0$) 
and in panel (b) a sample of bright galaxies 
($-20.0\,\ge\,M_{b_{\rm J}}-5\log_{10}\,h\,\ge\,-21.0$). 
Within each panel, early and late type galaxies, as distinguished 
by their spectral types, are plotted with 
different symbols; the positions of early-types are indicated by 
circles and the late-types are marked by stars. 
For clarity, we show only a three degree slice in declination cut 
from the SGP region and we have sparsely sampled the galaxies, so that the 
space densities of the two spectral classes are equal. 
In order to expand the scale of the plot, the range of redshifts shown is 
restricted, taking a subset of the full volume-limited sample in each case. 
(Note also that the redshift ranges differ between the two panels.)

A hierarchy of structures is readily apparent in these plots, ranging 
from isolated objects, to groups of a handful of galaxies and 
on through to rich clusters containing over a hundred members.  
It is interesting to see how structures are traced by galaxies in the 
different luminosity bins by comparing common structures between 
the two panels. For example, the prominent structure (possibly a 
supercluster of galaxies) seen at $\alpha\,\simeq\,0^{\rm h}$ and 
$z\,\simeq\,0.061$ is clearly visible in both panels. 
The same is true for the overdensity seen at $\alpha\,\simeq\,03^{\rm h}15'$ 
at $z\,\simeq\,0.068$. 

This is the first time that a large enough survey has been available, 
both in terms of the volume spanned and the number of measured redshifts, 
to allow a comparison of the clustering of galaxies of different 
spectral types and luminosities in representative volume-limited 
samples, without the complication of the strong radial gradient 
in number density seen in flux-limited samples.

It is apparent from a comparison of the distribution of the different 
spectral types in Fig.~\ref{fig:coneplot}(a), that the faint early-type 
galaxies tend to be grouped into structures on small scales whereas the 
faint late-types are more spread out.
One would therefore anticipate that the early-types should have a stronger 
clustering amplitude than the late-types, an expectation that is borne 
out in Section~\ref{sec:xi.s}.

\begin{figure}
\epsfig{file=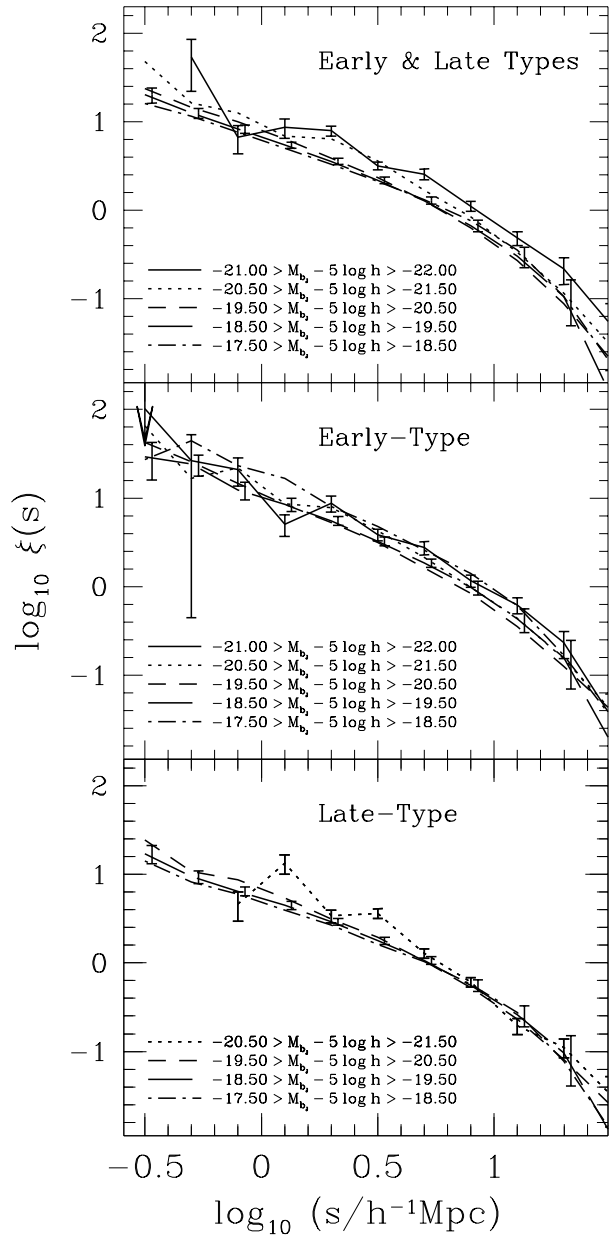,width=0.45\textwidth,clip=,bbllx=40,bblly=170,bburx=300,bbury=706}
\caption{
The spherically averaged redshift space correlation function of  
galaxies in disjoint absolute magnitude bins, as indicated by the 
key in each panel.
The panels show the results for different samples: 
the top panel shows the correlation functions for all galaxies 
that have been assigned a spectral type, the middle panel show the 
clustering of galaxies with $\eta<-1.4$ and the 
lower panel shows $\xi(s)$ for galaxies with $\eta>-1.4$. 
The error bars are obtained using 2dFGRS mock catalogues, 
as described in the text. For clarity, error bars are plotted 
only on the $-18.5\,\ge\,M_{b_{\rm J}}-5\log_{10}\,h\,\ge\,-19.5$ 
sample curve and for the brightest sample in each panel.}
\label{fig:xis}
\end{figure}

In Fig.~\ref{fig:coneplot}(b), the distinction 
between the distribution of the spectral types is less apparent. 
This is partly due to the greater importance of projection effects 
in the declination direction, as the cone extends to a greater redshift 
than in Fig.~\ref{fig:coneplot}(a).
However, close examination of the largest structures suggests that 
early-types are more abundant in them than late-types, again implying a
stronger clustering amplitude.

\subsection{$\xi(s)$ as function of luminosity and spectral type}
\label{sec:xi.s}

In Fig.~\ref{fig:xis}, we show the spherically averaged 
redshift space correlation function, $\xi(s)$, as a function 
of luminosity and spectral type. Results are shown for samples 
selected in bins of width one magnitude, as indicated by the 
legend in each panel. The top panel shows the correlation 
functions of all galaxies that have been assigned a spectral 
type, the middle panel shows results for 
galaxies classified as early-types ($\eta<-1.4$) and the bottom panel 
shows results for late-types ($\eta>-1.4$). 
Note that, at present, there are insufficient numbers of late-type 
galaxies to permit a reliable measurement of the correlation 
function for the brightest magnitude 
bin $-21.0\,\ge\,M_{b_{\rm J}}-5\log_{10}\,h\,\ge\,-22.0$. 

Several deductions can be made immediately from Fig.~\ref{fig:xis}.
In all cases, the redshift space correlation function is well described 
by a power-law {\it only} over a fairly limited range of scales. 
The correlation functions of early-type galaxies are somewhat steeper 
than those of late-types.  
However, the main difference is that the early-type galaxies have a 
stronger clustering amplitude than the late-type galaxies.  
The correlation length, defined here as the pair separation 
for which $\xi(s_{0})\,=\,1$,  varies for early-types 
from $s_{0}\,=\,7.1\,\pm\,0.7\,$\mpc\ for galaxies with absolute magnitudes 
around $M_{b_{\rm J}}-5\log_{10}\,h\,\sim\,-19.5$ 
to $s_{0}\,=\,8.9\,\pm\,0.7\,$\mpc\ for the brightest sample with 
$-21.0\,\ge\,M_{b_{\rm J}}-5\log_{10}\,h\,\ge\,-22.0$. 
The faintest early-types, with magnitudes 
$-17.5\,\ge\,M_{b_{\rm J}}-5\log_{10}\,h\,\ge\,-18.5$,   
display a clustering amplitude that is similar to that of 
the brightest early-types. However, the measurement of the correlation 
function for this faint sample is relatively noisy, as the volume in 
which galaxies are selected is small compared with the volumes used for 
brighter samples. 
The late-type galaxies show, by contrast, little change in clustering 
amplitude with increasing luminosity, with a redshift space correlation 
length of $s_{0}\,=\,5.6\,\pm\,0.6\,$\mpc. Only a slight steepening of the 
redshift space correlation function is apparent with increasing 
luminosity, until the brightest sample, which displays a modest increase 
in the redshift space correlation length. The correlation lengths of all 
our samples of early-type galaxies are larger than those of late-type 
galaxies.

\begin{figure*}
\epsfig{file=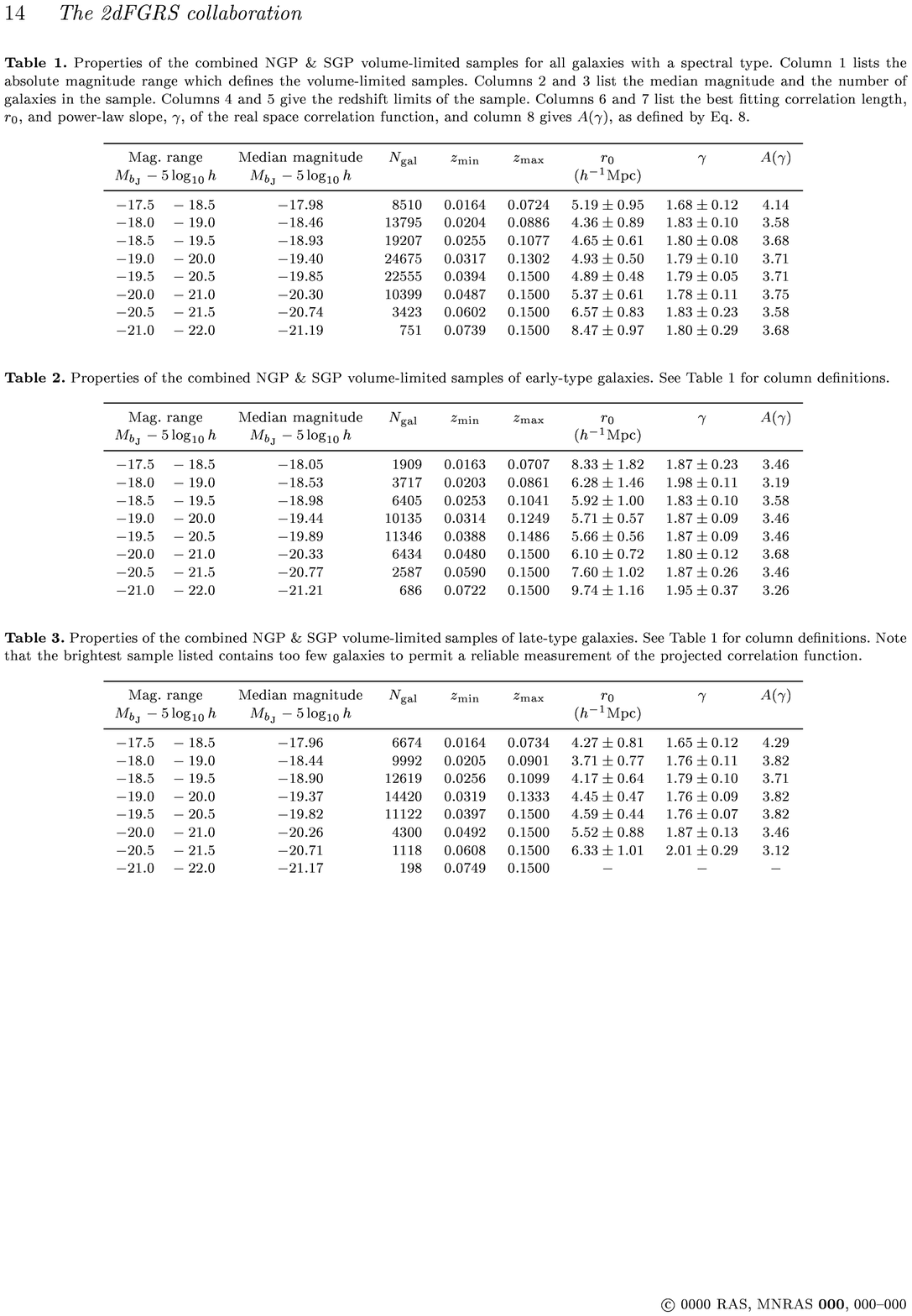,width=1.0\textwidth,clip=,bbllx=40,bblly=275,bburx=546,bbury=740}
\end{figure*}

\section{Clustering in real space}
\label{sec:res}

\subsection{Robustness of clustering results}
\label{sec:res.2mag}

The approach adopted to study the dependence of galaxy 
clustering on luminosity relies upon being able to compare correlation 
functions measured in different volumes. It is important to ensure   
that there are no systematic effects, such as significant sampling 
fluctuations, that could undermine such an analysis. 
In \paperone\, we demonstrated the robustness of this approach in two ways. 
First, we constructed a volume-limited sample defined using 
a broad magnitude range, that could be divided 
into co-spatial subsamples of galaxies in different luminosity bins, 
i.e. subsamples within the same volume and therefore subject to the same 
large-scale structure fluctuations. 
A clear increase in clustering amplitude was found for the 
brightest galaxies in the volume, establishing the dependence 
of clustering on galaxy luminosity (see Fig. 1a of Norberg \etal 2001). 
Secondly, we demonstrated that measuring the correlation 
function of galaxies in a fixed luminosity bin, but using samples taken from 
different volumes, gave consistent results 
(see Fig. 1b of Norberg \etal 2001). 

\begin{figure}
\epsfig{file=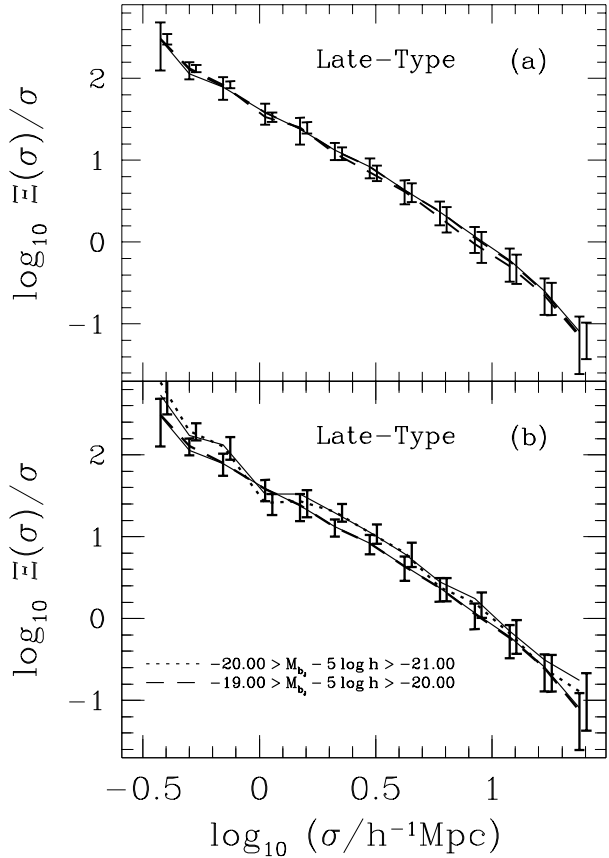,width=0.45\textwidth,clip=,bbllx=40,bblly=170,bburx=300,bbury=545}
\caption{
(a) The projected correlation function of late-type galaxies 
in a fixed absolute magnitude bin taken from different, almost 
independent volumes. We show the correlation function of galaxies with 
$-19.0\,\ge\,M_{b_{\rm J}}-5\log_{10}\,h\,\ge\,-20.0$ taken from volumes 
defined by  $-18.0\,\ge\,M_{b_{\rm J}}-5\log_{10}\,h\,\ge\,-20.0$  
and $-19.0\,\ge\,M_{b_{\rm J}}-5\log_{10}\,h\,\ge\,-21.0$ (both shown by 
heavy dashed lines). 
The thin solid line shows the estimate from the optimal sample for the 
$-19.0\,\ge\,M_{b_{\rm J}}-5\log_{10}\,h\,\ge\,-20.0$ magnitude bin. 
The different measurements are in almost perfect agreement.
(b) The projected correlation function measured for late-type galaxies 
in two {\it different} absolute magnitude bins taken from the {\it same\/} 
volume. The volume is defined by the magnitude range 
$-19.0\,\ge\,M_{b_{\rm J}}-5\log_{10}\,h\,\ge\,-21.0$. 
Within a fixed volume, there is clear evidence for an increase (albeit small) 
in the clustering amplitude with luminosity.
The two thin solid lines show estimates obtained from the corresponding 
optimal samples for the stated magnitude bins. 
In both panels the error bars come from the analysis 
of mock 2dFGRS catalogues.}
\label{fig:xr.test}
\end{figure}

In this section we repeat these tests. The motivation 
for this exercise is that the samples considered here 
contain fewer galaxies than those used by \paperone\, as only galaxies with $z<0.15$ 
are suitable for PCA spectral typing, and because the samples are more dilute 
as they have been selected on the basis of spectral type as well as 
luminosity.
In Fig.~\ref{fig:xr.test}(a) we plot the projected correlation function 
of late-type galaxies in a fixed absolute magnitude bin 
($-19.0\,\ge\,M_{b_{\rm J}}-5\log_{10}\,h\,\ge\,-20.0$), 
but measured for samples taken from volumes defined by different 
\zmin\ and \zmax.
The clustering results are in excellent agreement with one another. 
In Fig.~\ref{fig:xr.test}(b), we compare the projected correlation 
function of late-type galaxies in different absolute magnitude 
ranges but occupying the same volume.
A clear difference in the clustering amplitude is seen. 
We have also performed these tests for early-type galaxies and 
arrived at similar conclusions.

As an additional test, we also show in Fig.~\ref{fig:xr.test} the 
correlation function measured in what we refer to as the {\it optimal 
sample} for a given magnitude bin. The optimal sample contains the 
maximum number of galaxies for the specified magnitude bin.
The correlation functions of galaxies in optimal samples are shown 
by thin solid lines in both panels and are in excellent agreement 
with the other measurements shown.

\subsection{Projected correlation function}
\label{sec:res.xir}

\begin{figure}
\epsfig{file=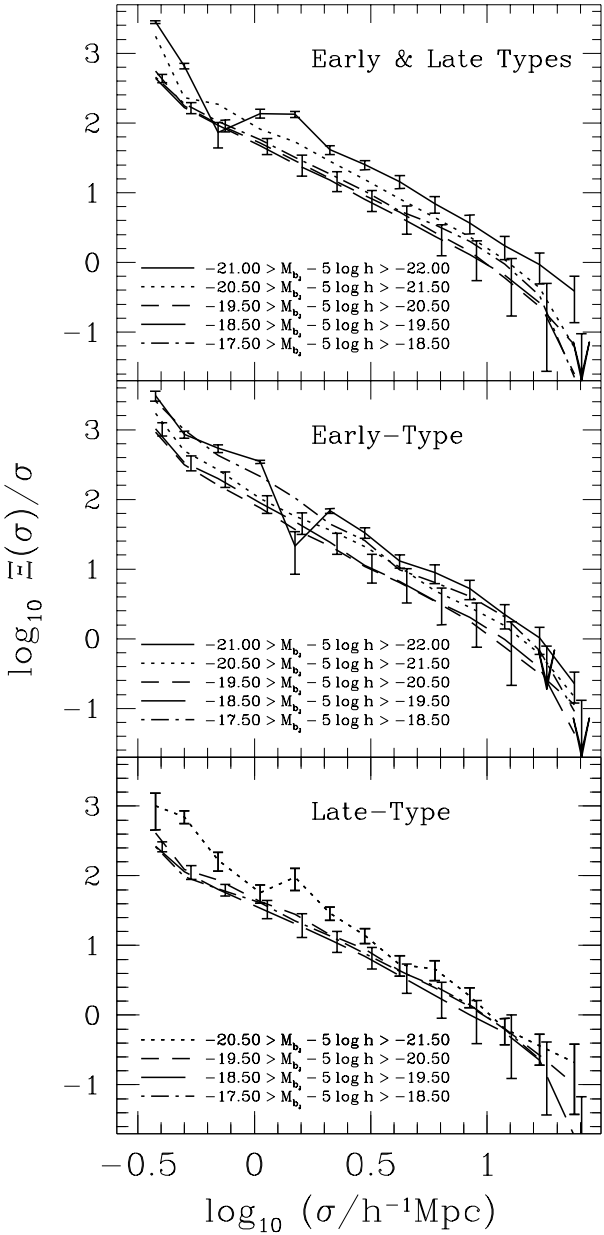,width=0.45\textwidth,clip=,bbllx=40,bblly=170,bburx=300,bbury=706}
\caption{
The projected galaxy correlation function for samples of different 
spectral type, split into one absolute magnitude wide bins.
The top panel shows the correlation function measured for all galaxies 
with an assigned spectral type.
The middle panel shows correlation functions for early-types and the bottom  
panel shows the results for late-types.
The absolute magnitude range of each sample is indicated in the legend 
on each panel.
The error bars are derived from the 2dFGRS mock catalogues and, 
for clarity, are only plotted on the correlation functions of the 
$-18.5\,\ge\,M_{b_{\rm J}}-5\log_{10}\,h\,\ge\,-19.5$ sample and of  
the brightest samples in each panel.
}
\label{fig:xr.tot}
\end{figure}

Fig.~\ref{fig:xr.tot} shows how the real space clustering 
of galaxies of different spectral type depends on luminosity. 
We use the optimal sample for each magnitude bin, i.e. the volume-limited 
sample with the maximum possible number of galaxies, the properties 
of which are listed in Tables~\ref{tab:eta0} (early \& late types 
together),~\ref{tab:eta1} (early-types only) 
and~\ref{tab:eta6} (late-types only).

The top panel of Fig.~\ref{fig:xr.tot} confirms the results 
found by \paperone, namely that the clustering strength of the full 
sample increases slowly with increasing luminosity for galaxies 
fainter than \Mstar, and then shows a clear, strong increase for 
galaxies brighter than \Mstar. (We take \Mstar to be 
$M_{b_{\rm J}}-5\log_{10}\,h\,\simeq\,-19.7$, following 
Folkes \etal 1999 and Norberg \etal 2002.)
Furthermore, the projected correlation functions are well described by 
a power law with a slope that is independent of luminosity.
The middle panel of Fig.~\ref{fig:xr.tot} shows the projected correlation 
function of early-type galaxies for different absolute magnitude ranges. 
The clustering amplitude displays a non-monotonic behaviour, with the 
faintest sample having almost the same clustering strength as the 
brightest sample. The significance of this result for the 
faintest galaxies will be discussed further in the next section. 
Early-type galaxies with $M_{b_{\rm J}}-5\log_{10}\,h\,\simeq\,-19.5$, display 
weaker clustering than the faint and bright samples.
The bottom panel of Fig.~\ref{fig:xr.tot} shows the real space 
clustering of late-type galaxies as a function of luminosity. 
In this case, the trend of clustering strength with luminosity is much 
simpler.
There is an increase in clustering amplitude with luminosity, 
and also some evidence that the projected correlation function of the 
brightest subset is steeper than that of the other late-type samples.
In general, for the luminosity ranges for which a comparison can be made, 
the clustering strength of early-type galaxies is always stronger than 
that of late-types.

\begin{figure}
\epsfig{file=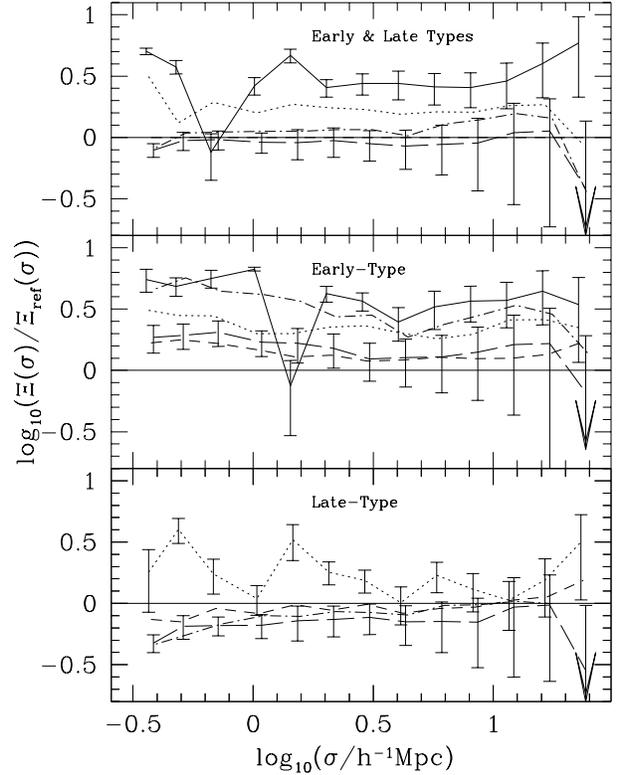,width=0.45\textwidth,clip=,bbllx=20,bblly=155,bburx=440,bbury=706}
\caption{
The ratio of the projected correlation function measured for 
galaxies selected by luminosity and spectral type to the projected 
correlation function of a reference sample. The reference sample 
consists of all galaxies with an assigned spectral type that lie within 
the magnitude range $-19.5\,\ge\,M_{b_{\rm J}}-5\log_{10}\,h\,\ge\,-20.5$.
The top panel shows the ratios for different luminosity bins for 
all galaxies with a spectral type, the middle panel shows the ratios 
obtained for early-types and the bottom panel shows the ratios 
for late-types. 
The same line styles plotted in Fig.~\ref{fig:xr.tot} are used to 
indicate different luminosities.
The error bars, which take into account the correlation between the 
samples (but not between the bins), are from the mock 2dFGRS 
catalogues, and for clarity, are only plotted on two curves in each panel: 
the ratio for the sample with 
$-18.5\,\ge\,M_{b_{\rm J}}-5\log_{10}\,h\,\ge\,-19.5$ and the 
ratio for the brightest sample in each panel. 
}
\label{fig:ratio}
\end{figure}

The comparison of the correlation functions of different 
samples is made simpler if we divide the curves plotted in 
Fig.~\ref{fig:xr.tot} by a fiducial correlation function. 
As a reference sample we choose 
all galaxies that 
have been assigned a spectral type, with absolute magnitudes in the 
range $-19.5\,\ge\,M_{b_{\rm J}}-5\log_{10}\,h\,\ge\,-20.5$ 
(the short-dashed line in the top panel of Fig.~\ref{fig:xr.tot}).
In Fig.~\ref{fig:ratio}, we plot the ratio of the correlation functions 
shown in the panels of Fig.~\ref{fig:xr.tot}, to the reference correlation 
function, with error bars obtained from the mock catalogues.
The trends reported above for the variation of clustering strength 
with luminosity and spectral type are now clearly visible 
(see~Fig.~\ref{fig:ratio}), particularly the difference in clustering 
amplitude between early-types and late-types.
In the upper and lower panels, the ratios of correlation functions 
are essentially independent of $\sigma$, 
indicating that a single power-law slope is 
a good description over the range of scales plotted. 
The one exception is the brightest sample of late-type galaxies, 
which shows some evidence for a steeper power-law.
In the middle panel, the ratios for early-type galaxies   
show tentative evidence for a slight steepening of the correlation 
function at small pair separations, $\sigma<2h^{-1}\,$Mpc, which 
is most pronounced for the brightest sample.

\subsection{Real space correlation length}
\label{sec:xi.r0}

\begin{figure}
\epsfig{file=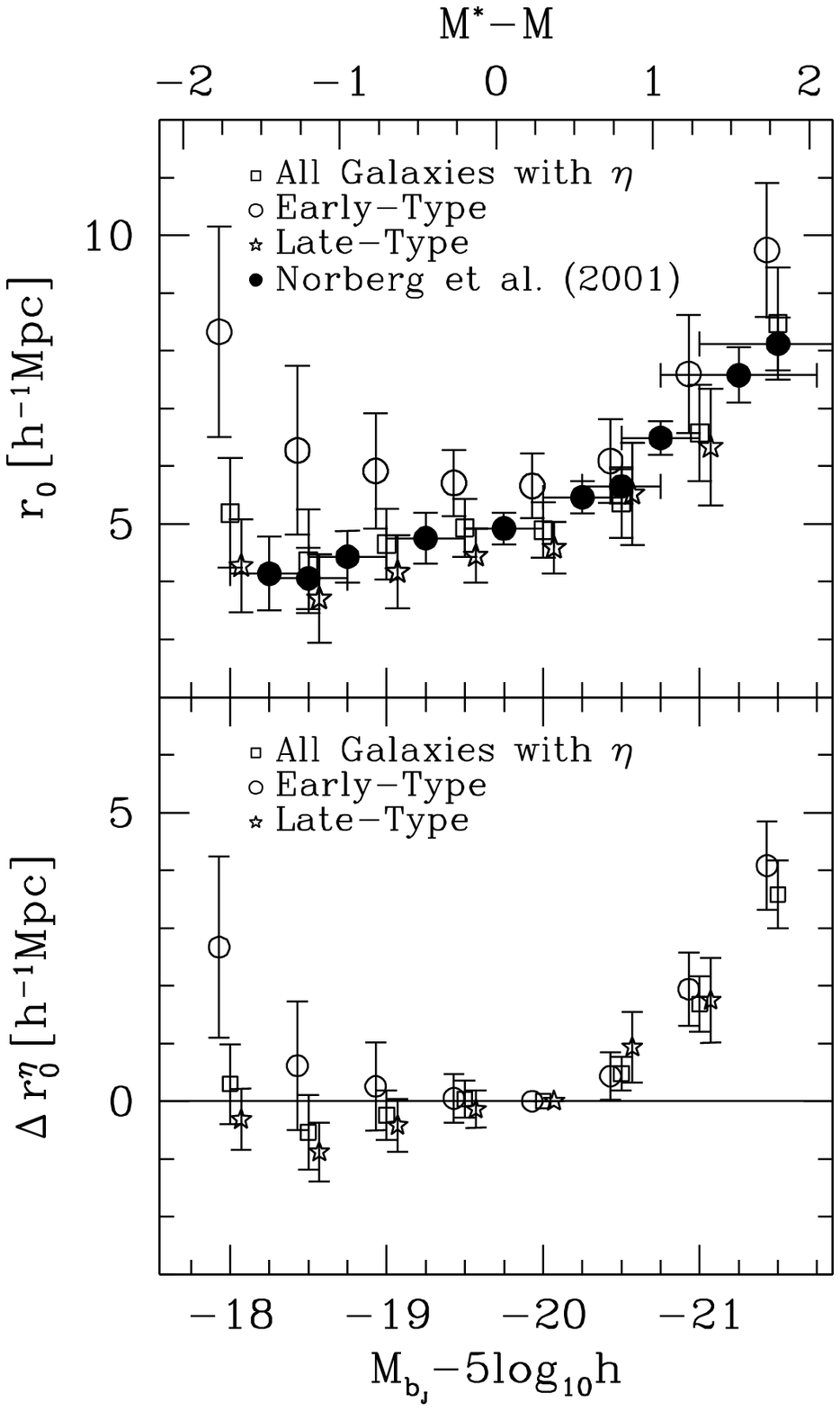,width=0.47\textwidth,clip=,bbllx=60,bblly=160,bburx=375,bbury=685}
\caption{Top panel: the real space correlation length, $r_0$, as a 
function of the absolute magnitude of the sample, for galaxies 
of different spectral type. The stars show 
the $r_0$ values fitted to the projected correlation 
function of late-type galaxies and the open circles show 
the $r_0$ values for early-types. 
The squares show $r_0$ for the full sample with spectral types.
The latter results are in excellent agreement with 
those obtained from the larger sample analysed by \paperone, 
which are plotted as filled circles.
The horizontal bars on the filled circles show the magnitude range 
used to define the volume-limited samples.
Lower panel: the difference $\Delta r_0^\eta\,=\,r_{0}^\eta({\rm M_{b_{\rm J}}})-r_{0}^\eta({\rm M_{b_{\rm J}}^{\rm ref}})$ for each spectral type. 
This quantity gives the significance of the variation of the correlation 
length with luminosity for each spectral type with respect to a 
reference sample, chosen to be the sample defined 
by $-19.5\,\ge\,M_{b_{\rm J}}-5\log_{10}\,h\,\ge\,-20.5$. The error bars 
take into account the correlation between the two samples.}
\label{fig:r0}
\end{figure}

In the previous subsection, we demonstrated that the projected 
correlation functions of galaxies in the 2dFGRS have 
a power-law form with a slope that varies little as the sample 
selection is changed, particularly for pair separations in the range 
$2.0\,\le\,\sigma/(\,h^{-1}\,{\rm Mpc})\,\le\,15.0$.
To summarize the trends in clustering strength found when varying 
the spectral type and luminosity of the sample, 
we fit a power-law over this range of scales. 
We follow the approach, based on Eq.~\ref{eq:pow}, used by \paperone, who 
performed a $\chi^{2}$ 
minimisation to extract the best fitting values of the parameters 
in the power-law model for the real space correlation function: 
the correlation length, $r_0$, and the power law slope, $\gamma$.
As pointed out by \paperone, a simple $\chi^2$ approach does not  
give reliable estimates of the errors on the fitted parameters 
because of the correlation between the estimates of the 
correlation function at differing pair separations. 
We therefore use the mock 2dFGRS catalogues to estimate the errors on the 
fitted parameters in the following manner. 
The best fitting values of $r_0$ and $\gamma$ are found for 
each mock individually, using the $\chi^2$ analysis.
The estimated error is then taken to be the fractional {\it rms}\ scatter 
in the fitted parameters over the ensemble of mock catalogues. 
The best fitting parameters for each 2dFGRS sample are listed in 
Tables~\ref{tab:eta0} (early \& late types together),~\ref{tab:eta1} 
(early-types only) and~\ref{tab:eta6} (late-types only). The results 
for the correlation length are plotted in the top panel of Fig.~\ref{fig:r0}. 

The correlation lengths estimated for the full sample with assigned spectral 
types (shown by the open squares in Fig.~\ref{fig:r0}) are in excellent 
agreement with the results of \paperone\ (shown by the filled circles). 
The bright samples constructed by \paperone\ have \zmax\ values 
in excess of the limit of \zmax$=0.15$ enforced upon the samples analysed 
in this paper by the PCA. 
The bright samples used in this paper therefore cover smaller volumes 
than those used by Norberg \etal and so the error bars are 
substantially larger.
A cursory inspection of Fig.~\ref{fig:r0} would give the misleading 
impression that we find weaker evidence for an increase in correlation 
length with luminosity. 
It is important to examine this plot in conjunction 
with Tables~\ref{tab:eta0} to~\ref{tab:eta6}, which reveals that there 
is significant overlap in the volumes defined by the four brightest 
magnitude slices, because of the common \zmax$=0.15$ limit.

In this case, the error bars inferred directly from the mocks do not 
take into account that our samples are correlated. The errors fully 
incorporate cosmic variance, {\it i.e.} the variance in clustering 
signal expected when sampling a given volume placed at different, independent  
locations in the Universe. The volumes containing the four brightest 
samples listed in Table~\ref{tab:eta1} contain long-wavelength 
fluctuations in common and so clustering measurements from these 
different volumes are subject to a certain degree of coherency. 
The clearest way to show this is by calculating the 
difference between the correlation lengths fitted to two samples. The error on the 
difference, derived using the mock catalogues, has two components: 
the first comes from adding the individual errors in 
quadrature; the second is from the correlation of the 
samples. For correlated samples, this second term is negative 
and, therefore lowers the estimated error on the difference. This is 
precisely what is seen in the lower panel of Fig.~\ref{fig:r0}, where 
we plot $\Delta\,r_0^\eta\,=\,r_{0}^\eta({\rm M_{b_{\rm J}}})
- r_{0}^\eta({\rm M_{b_{\rm J}}^{\rm ref}})$ 
for each spectral type as a function of absolute magnitude, with error 
bars taking into account the correlation of the samples. For all 
samples brighter than \Mstar, the increase in clustering length 
with luminosity is clear. 

There is a suggestion, in the top panel of Fig.~\ref{fig:r0}, of a 
non-monotonic dependence of the correlation length 
on luminosity for early-type galaxies. The evidence for this 
behaviour is less apparent on the $\Delta\,r_0^\eta$ panel, 
where a difference in $r_0$ for the faintest sample is seen at less than 
the 2$\sigma$ level. These volumes are small compared to those defining 
the brighter samples. Furthermore, when analyzing the faintest sample 
for SGP and NGP separately, our results are somewhat sensitive to the presence of 
single structures in each region (at $\alpha\,\simeq\,0^{\rm h}$ 
and $z\,\simeq\,0.061$ for the SGP, as shown in 
Fig.~\ref{fig:coneplot}(a)). Thus we conclude that the upturn 
at faint magnitudes in the correlation length of early-types is 
not significant.

The projected correlation function of early-type galaxies brighter than 
\Mstar\ is well fitted by a power-law real space correlation 
function, with a virtually constant slope of $\gamma\,\simeq\,1.87\,$
and a correlation length which increases with luminosity, from 
$r_0\,=\,5.7\,\pm\,0.6\,$\mpc\ for \Mstar\ galaxies to
$r_0\,=\,9.7\,\pm\,1.2\,$\mpc\ for brighter galaxies 
($M_{b_{\rm J}}-5\log_{10}\,h\,\simeq\,-21.2$). 
This represents an increase in clustering strength by 
a factor of $2.7$, as seen in Fig.~\ref{fig:ratio}.
The projected correlation functions of late-type galaxies are also 
consistent with a power-law in real space, with 
an essentially constant slope. There is a very weak trend for $\gamma$ 
to increase with luminosity, although at little 
more than the $1 \sigma$ level.
Ignoring this effect, the fitted slope of the late-type correlation function 
is $\gamma\,\simeq\,1.76\,$.
The  correlation length increases with luminosity from a value of 
$r_0\,=\,3.7\,\pm\,0.8\,$\mpc\ for faint galaxies 
($M_{b_{\rm J}}-5\log_{10}\,h\,\simeq\,-18.4$) to
$r_0\,=\,6.3\,\pm\,1.0\,$\mpc\ for bright galaxies 
($M_{b_{\rm J}}-5\log_{10}\,h\,\simeq\,-20.7$), 
a factor of $2.5$ increase in clustering strength.
It should be possible to extend the analysis for late-type galaxies 
beyond $M_{b_{\rm J}}-5\log_{10}\,h\,\simeq\,-21$ when the 2dFGRS is complete.

The top panel of Fig.~\ref{fig:r0} confirms our earlier conclusion 
that the clustering 
of early-type galaxies is stronger than that of late-type galaxies. 
At \Mstar, early-types typically have a real space clustering 
amplitude that is $1.5-1.7$ times greater than that of late-types.

\section{Discussion and conclusions}
\label{sec:conc}

\begin{figure}
\epsfig{file=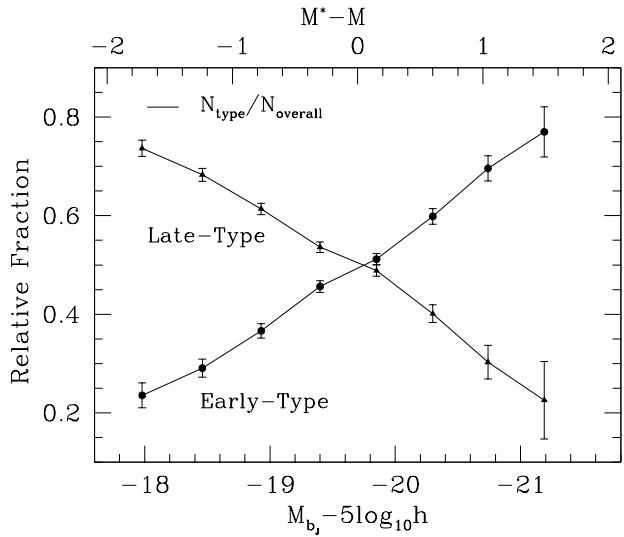,width=0.47\textwidth,clip=,bbllx=50,bblly=205,bburx=550,bbury=640}
\caption{
The fraction of galaxies in the two broad spectral classes, early-type and 
late-type, as a function of absolute magnitude. The fractions are derived 
from the volume-limited samples listed in Tables 1, 2 and 3. 
The error bars show the Poisson errors on the fractions.
}
\label{fig:ndens}
\end{figure}

We have used the 2dFGRS to study the dependence of clustering on 
spectral type for samples spanning a factor of twenty in galaxy luminosity.
The only previous attempt at a bivariate luminosity-morphology/spectral 
type analysis of galaxy clustering was performed by Loveday \etal (1995) 
using the Stromlo-APM redshift survey. 
They were able to probe only a relatively narrow range in 
luminosity around $L^{\star}$, which is more readily apparent if 
one considers the median magnitude of each of their magnitude 
bins (see Fig. 3b of Norberg \etal 2001). 
The scatter between spectral and morphological types illustrated 
in Fig.~\ref{fig:jon.eta} precludes a more detailed comparison of 
our results with those of earlier studies, based on morphological 
classifications.

In Norberg \etal (2001), we used the 2dFGRS to make a precise 
measurement of the dependence of galaxy clustering on 
luminosity.
The clustering amplitude was found to scale linearly with 
luminosity.
One of the aims of the present paper is to identify the 
phenomena that drive this relation. 
In particular, there are two distinct hypotheses that we wish 
to test. The first is that there is a general trend for 
clustering strength to increase with luminosity, regardless of 
the spectral type of the galaxy.
The second is that different types of galaxies have 
different clustering strengths, which may 
vary relatively little with luminosity, but a variation of the 
mix of galaxy types with luminosity results in a dependence of the 
clustering strength on luminosity.

Madgwick \etal (2002) estimated the luminosity function of 2dFGRS  
galaxies for different spectral classes, and found that, in going 
from early-type to late-types, the slope of the faint end of the 
luminosity function becomes steeper while the characteristic 
luminosity becomes fainter.
Another representation of the variation of the luminosity function  
with spectral class is shown in Fig.~\ref{fig:ndens}, where we plot the 
fraction of early and late type galaxies in absolute magnitude bins. 
The plotted fractions are derived from the volume-limited samples listed 
in Tables~1,~2 and~3. 
The mix of spectral types changes dramatically with 
luminosity; faint samples are dominated by late-types, whereas 
early-types are the most common galaxies in bright samples. 
Similar trends were found for galaxies labelled by morphological type 
in the SSRS2 survey by Marzke \etal (1998).

\begin{figure}
\epsfig{file=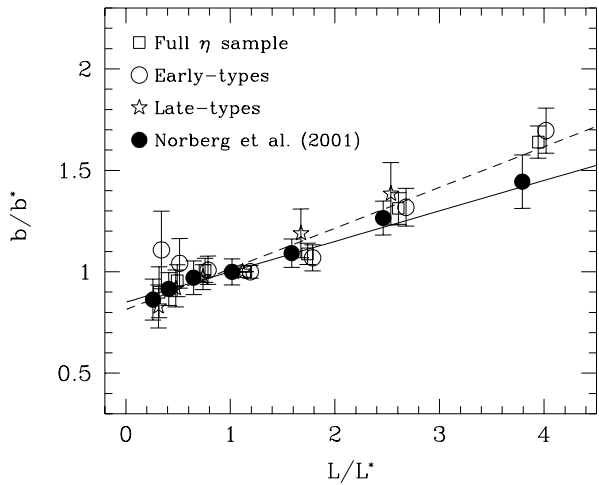,width=0.47\textwidth,clip=,bbllx=47,bblly=225,bburx=560,bbury=672}
\caption{
The relative bias (on the scale of $r = 4.89$\mpc) of the 
different spectral classes as a function of 
luminosity, as indicated by the key. 
The definition of relative bias is given in the text. The 
reference sample covers the absolute magnitude range 
$-19.5\,\ge\,M_{b_{\rm J}}-5\log_{10}\,h\,\ge\,-20.5$.
The fiducial luminosity, 
$L^{\star}$, is taken to be $M_{b_{\rm J}}-5\log_{10}\,h\,=-19.7$.
The solid line shows the fit to the results of \paperone, 
$b/b^{\star} = 0.85 + 0.15L/L^{\star}$, whereas the dashed line 
shows the fit to the data analyzed here. 
All the error bars plotted take into account the correlation 
between the various samples.
}
\label{fig:bl}
\end{figure}

We find that the change in the mix of spectral types with luminosity is 
not the main cause for the increase in the clustering strength 
of the full sample with luminosity. To support this assertion, 
we plot in Fig.~\ref{fig:bl} the variation of clustering strength with 
luminosity normalized, for {\it each} spectral class, to the clustering 
strength of a fiducial sample of \Mstar\ galaxies, {\it i.e.} the sample 
which covers the magnitude range 
$-19.5\,\ge\,M_{b_{\rm J}}-5\log_{10}\,h\,\ge\,-20.5$. 
For a galaxy sample with best fitting correlation function parameters  
$r^{i}_{0}$ and $\gamma^{i}$, we define the relative bias with respect to 
the \Mstar\ sample of the same type by 
\begin{equation} 
\frac{b^{i}}{b^{\star}}\Big|_{\eta}(r) = 
\sqrt{\frac{ (r^{i}_{0})^{\gamma_{i}}}{r^{\gamma}_{0}} 
r^{\gamma-\gamma_{i}}}\Big|_{\eta},
\label{eq:bias}
\end{equation}
where $r_{0}$ and $\gamma$ are the best fitting power-law parameters for the 
fiducial sample. In Fig.~\ref{fig:bl}, we plot the relative bias 
evaluated at a fixed scale, $r = 4.89$\mpc, which is the correlation 
length of the reference sample for all \etapar-classified galaxies.
A scale dependence in Eq.~\ref{eq:bias} arises if the slopes of the 
real space correlation functions are different for the galaxy samples 
being compared. In practice, the term $r^{\gamma-\gamma_{i}}$ is 
close to unity for the samples considered.
The dashed line shows a fit to the bias relation defined by the open 
symbols.
The solid line shows the effective bias relation obtained by \paperone, 
which is defined in a slightly different way to the effective bias 
computed here. 

From Fig.~\ref{fig:bl}, we see that 
the trend of increasing clustering strength with luminosity in both 
spectral classes is very similar for galaxies brighter than $L>0.5L^{\star}$.
At the brightest luminosity, corresponding to $\sim\,4 L^{\star}$, 
the clustering amplitude is a factor of $2-2.5$ times greater 
than at $L^{\star}$. This increase is much larger than the offset 
in the relative bias factors of early and late types at any given 
luminosity. 
We conclude that the change in correlation length with absolute 
magnitude found by Norberg \etal (2001) is primarily a luminosity effect 
rather than a reflection of the change in the mix of spectral 
types with luminosity. 

Benson \etal (2001) showed that a dependence of clustering strength 
on luminosity is expected in hierarchical clustering cold dark matter 
universes because of the preferential formation of the brightest 
galaxies in the most massive, strongly clustered dark halos. The 
close connection between the spectral characteristics of galaxies 
and their clustering properties discussed in this paper provides 
further evidence that the galaxy type is also related to the mass 
of the halo in which galaxy forms.

\section*{Acknowledgments}
The 2dFGRS is being carried out using the 2 degree field facility on the 
3.9m Anglo\-Australian Telescope (AAT). We thank all those involved in 
the smooth running and continued success of the 2dF and the AAT. 
We thank the referee, Dr. J. Loveday, for producing a speedy and 
helpful report. PN is supported by the Swiss National Science 
Foundation and an ORS award, and CMB acknowledges the receipt of a 
Royal Society University Research Fellowship. 
This work was supported in part by a PPARC rolling grant at Durham.

\begin{table*}
\caption{
Properties of the combined NGP \& SGP volume-limited samples 
for all galaxies with a spectral type. Column 1 lists the absolute 
magnitude range which defines the volume-limited samples. Columns 
2 and 3 list the median magnitude and the number 
of galaxies in the sample.  
Columns 4 and 5 give the redshift limits of the sample. 
Columns 6 and 7 list the best fitting correlation length, 
$r_{0}$, and power-law slope, $\gamma$, of the real space 
correlation function, and
column 8 gives $A(\gamma)$, as defined by Eq.~\ref{eq:pow}. 
}
\begin{tabular}{cccccccc}   
\hline
Mag. range           & Median magnitude   &  $N_{\rm gal}$ & $z_{\rm min}$   & $z_{\rm max}$ & $r_{0}$         & $\gamma$ & $A(\gamma)$ \\ 
$M_{b_{\rm J}}-5\log_{10} h$  & $M_{b_{\rm J}}-5\log_{10} h$       &            &                 &               & (\mpc)  &     &  \\ \hline 


$-17.5 \quad  -18.5$ & $-17.98$            &  \phantom{0}8510      &  0.0164         &   0.0724       &	 $5.19 \pm 0.95$ & $1.68 \pm 0.12$  & 4.14  \\ 

$-18.0 \quad  -19.0$ & $-18.46$            &            13795      &  0.0204         &   0.0886       &  $4.36 \pm 0.89$ & $1.83 \pm 0.10$  & 3.58  \\ 

$-18.5 \quad  -19.5$ & $-18.93$            &            19207      &  0.0255         &   0.1077       &  $4.65 \pm 0.61$ & $1.80 \pm 0.08$  & 3.68  \\ 

$-19.0 \quad  -20.0$ & $-19.40$            &            24675      &  0.0317         &   0.1302       &  $4.93 \pm 0.50$ & $1.79 \pm 0.10$  & 3.71  \\ 

$-19.5 \quad  -20.5$ & $-19.85$            &            22555      &  0.0394         &   0.1500       &  $4.89 \pm 0.48$ & $1.79 \pm 0.05$  & 3.71  \\ 

$-20.0 \quad  -21.0$ & $-20.30$            &            10399      &  0.0487         &   0.1500       &  $5.37 \pm 0.61$ & $1.78 \pm 0.11$  & 3.75  \\ 

$-20.5 \quad  -21.5$ & $-20.74$            &  \phantom{0}3423      &  0.0602         &   0.1500       &  $6.57 \pm 0.83$ & $1.83 \pm 0.23$  & 3.58  \\ 

$-21.0 \quad  -22.0$ & $-21.19$            &  \phantom{00}751      &  0.0739         &   0.1500       &  $8.47 \pm 0.97$ & $1.80 \pm 0.29$  & 3.68  \\ 

\end{tabular}
\label{tab:eta0}
\end{table*}

\begin{table*}
\caption{
Properties of the combined NGP \& SGP volume-limited 
samples of early-type galaxies. See Table~\ref{tab:eta0} 
for column definitions.
}
\begin{tabular}{cccccccc}   
\hline
Mag. range           & Median magnitude   &  $N_{\rm gal}$ & $z_{\rm min}$   & $z_{\rm max}$ & $r_{0}$         & $\gamma$ & $A(\gamma)$ \\ 
$M_{b_{\rm J}}-5\log_{10} h$  & $M_{b_{\rm J}}-5\log_{10} h$       &            &                 &               & (\mpc)  &     &    \\ \hline 


$-17.5 \quad  -18.5$ & $-18.05$            &  \phantom{0}1909       &  0.0163          &   0.0707       & $8.33 \pm 1.82$ & $1.87 \pm 0.23$  & 3.46   \\ 

$-18.0 \quad  -19.0$ & $-18.53$            &  \phantom{0}3717       &  0.0203          &   0.0861       & $6.28 \pm 1.46$ & $1.98 \pm 0.11$  & 3.19   \\ 

$-18.5 \quad  -19.5$ & $-18.98$            &  \phantom{0}6405       &  0.0253          &   0.1041       & $5.92 \pm 1.00$ & $1.83 \pm 0.10$  & 3.58   \\ 

$-19.0 \quad  -20.0$ & $-19.44$            &            10135       &  0.0314          &   0.1249       & $5.71 \pm 0.57$ & $1.87 \pm 0.09$  & 3.46   \\ 

$-19.5 \quad  -20.5$ & $-19.89$            &            11346       &  0.0388          &   0.1486       & $5.66 \pm 0.56$ & $1.87 \pm 0.09$  & 3.46   \\ 

$-20.0 \quad  -21.0$ & $-20.33$            &  \phantom{0}6434       &  0.0480          &   0.1500       & $6.10 \pm 0.72$ & $1.80 \pm 0.12$  & 3.68   \\ 

$-20.5 \quad  -21.5$ & $-20.77$            &  \phantom{0}2587       &  0.0590          &   0.1500       & $7.60 \pm 1.02$ & $1.87 \pm 0.26$  & 3.46   \\ 

$-21.0 \quad  -22.0$ & $-21.21$            &   \phantom{00}686      &  0.0722          &   0.1500       & $9.74 \pm 1.16$ & $1.95 \pm 0.37$  & 3.26   \\ 

\end{tabular}
\label{tab:eta1}
\end{table*}

\begin{table*}
\caption{
Properties of the combined NGP \& SGP volume-limited 
samples of late-type galaxies.
See Table~\ref{tab:eta0} for column definitions.
Note that the brightest sample listed contains too few galaxies 
to permit a reliable measurement of the projected correlation function.
}
\begin{tabular}{ccccccccc}   
\hline
Mag. range           & Median magnitude   &  $N_{\rm gal}$ & $z_{\rm min}$   & $z_{\rm max}$ & $r_{0}$         & $\gamma$ & $A(\gamma)$ \\ 
$M_{b_{\rm J}}-5\log_{10} h$  & $M_{b_{\rm J}}-5\log_{10} h$       &            &                 &               & (\mpc)  &     &   \\ \hline 


$-17.5 \quad  -18.5$ & $-17.96$            &  \phantom{0}6674      &  0.0164          &   0.0734       & $4.27 \pm 0.81$ & $1.65 \pm 0.12$  & 4.29   \\ 

$-18.0 \quad  -19.0$ & $-18.44$            &  \phantom{0}9992      &  0.0205          &   0.0901       & $3.71 \pm 0.77$ & $1.76 \pm 0.11$  & 3.82   \\ 

$-18.5 \quad  -19.5$ & $-18.90$            &            12619      &  0.0256          &   0.1099       & $4.17 \pm 0.64$ & $1.79 \pm 0.10$  & 3.71   \\ 

$-19.0 \quad  -20.0$ & $-19.37$            &            14420      &  0.0319          &   0.1333       & $4.45 \pm 0.47$ & $1.76 \pm 0.09$  & 3.82   \\ 

$-19.5 \quad  -20.5$ & $-19.82$            &            11122      &  0.0397          &   0.1500       & $4.59 \pm 0.44$ & $1.76 \pm 0.07$  & 3.82   \\ 

$-20.0 \quad  -21.0$ & $-20.26$            &  \phantom{0}4300      &  0.0492          &   0.1500       & $5.52 \pm 0.88$ & $1.87 \pm 0.13$  & 3.46   \\ 

$-20.5 \quad  -21.5$ & $-20.71$            &  \phantom{0}1118      &  0.0608          &   0.1500       & $6.33 \pm 1.01$ & $2.01 \pm 0.29$  & 3.12   \\ 

$-21.0 \quad  -22.0$ & $-21.17$            &  \phantom{00}198      &  0.0749          &   0.1500       &	$-$		 &	$-$	   &	$-$   \\ 

\end{tabular}
\label{tab:eta6}
\end{table*}

\end{document}